\documentclass[prd,aps,floats,preprintnumbers,preprint]{revtex4}

\usepackage{graphicx}
\usepackage{multirow}
 
\newcommand{\be}{\begin{equation}}
\newcommand{\ee}{\end{equation}}
\newcommand{\bea}{\begin{eqnarray}}
\newcommand{\eea}{\end{eqnarray}}

\begin{document}
\setlength{\unitlength}{1mm}

\title{A new method to determine large scale structure from the luminosity distance}

\author{Antonio Enea Romano$^{1,2,3,5}$}
\author{Hsu-Wen Chiang$^{1,2,3}$}
\author{Pisin Chen$^{1,2,3,4}$}
\affiliation{
${}^1$Department of Physics, National Taiwan University, Taipei 10617, Taiwan, R.O.C.\\
${}^2$Leung Center for Cosmology and Particle Astrophysics, National Taiwan University, Taipei 10617, Taiwan, R.O.C.\\
${}^3$Graduate Institute of Astrophysics, National Taiwan University, Taipei 10617, Taiwan, R.O.C.\\
${}^4$Kavli Institute for Particle Astrophysics and Cosmology, SLAC National Accelerator Laboratory, Menlo Park, CA 94025, U.S.A. \\
${}^{5}$Department of Physics, McGill University, Montr\'eal, QC H3A 2T8, Canada \\
}



\begin{abstract}
The luminosity distance can be used to determine the properties of large scale structure around the observer. To this purpose we develop a new inversion method to map luminosity distance to a LTB metric based on the use of the exact analytical solution for Einstein equations. The main advantages of this approach are an improved numerical accuracy and stability, an exact analytical setting of the initial conditions for the differential equations which need to be solved and the validity for any sign of the functions determining the LTB geometry. Given the fully analytical form of the differential equations, this method  also simplifies the calculation of the red-shift expansion around the apparent horizon point where the numerical solution becomes unstable.
We test the method by inverting the supernovae Ia luminosity distance function corresponding to the the best fit $\Lambda CDM$  model. We find that only a limited range of initial conditions is compatible with observations, or a transition from red to blue shift can occur at relatively low redshift. 
Despite LTB solutions without a cosmological constant have been shown not to be compatible with all different set of available observational data, those studies normally fit data  assuming a special functional ansatz for the inhomogeneity profile, which often depend only on few parameters. Inversion methods on the contrary are able to fully explore the freedom in fixing the functions which determine a LTB solution. For this reason this inversion method could be applied to  explore more exhaustively  the compatibility with observations. Another important possible application is not about LTB solutions as cosmological models, but rather as tools to study the effects on the observations made by a generic observer located in an inhomogeneous region of the Universe where a fully non perturbative treatment involving exact solutions of Einstein equations is required.

\end{abstract}

\maketitle
\section{Introduction}

When different cosmological data \cite{Riess:1998cb,Perlmutter:1998np,Tonry:2003zg,Knop:2003iy,Barris:2003dq,Riess:2004nr,Bennett:2003bz,Spergel:2006hy} are interpreted using FLRW models a dominant dark energy must be introduced.
Since the nature of dark energy is not well understood, there has been some efforts to look for alternative explanations based on relaxing the hypothesis of large scale homogeneity. It is well known that inhomogeneous matter dominated models can fit some of the available observations \cite{Romano:2006yc,Chung:2006xh,Yoo:2008su,Romano:2009ej,Alexander:2007xx,Alnes:2005rw,GarciaBellido:2008nz,GarciaBellido:2008gd,GarciaBellido:2008yq,Romano:2012ks,February:2009pv,Romano:2009qx,Clarkson:2007bc,Ishibashi:2005sj,Bolejko:2011ys,Romano:2012yq,Clifton:2009kx,Romano:2007zz,Balcerzak:2013kha,Zibin:2011ma,Bull:2011wi,Romano:2009mr},
 and different methods have been developed to solve the inversion problem (IP) to map a given observed luminosity distance function $D_L(z)$ to the corresponding inhomogeneous metric. In this paper we will study  the case of a radially inhomogeneous spherically symmetric metric pressureless solution, described by a Lemaitre-Tolman-Bondi (LTB) solution, assuming a central location of the observer. This is an open violation of the Copernican principle, but since this is more a philosophical principle than a fully observationally established fact, it is worth investigation this type of cosmological model. Previous solutions to the inversion problems \cite{Chung:2006xh,Celerier:2009sv} were based on the solution of the radial light cone geodesics using a different system of coordinates, but they all required a numerical integration of the background Einstein's equations, while in this paper we derive a fully analytical method for the solution of the IP, based on the use of the exact analytical solution.
The use of galaxy number counts \cite{Romano:2007zz} has also been proposed to distinguish 
between inhomogeneous models  and $\Lambda$CDM  
 using both both analytical  and 
numerical approaches \cite{Romano:2009qx,Romano:2009ej,Celerier:2009sv}.

There has also been a considerable interest on the  effects of large scale inhomogeneities in presence of dark energy \cite{Romano:2010nc,Romano:2012gk,Romano:2012kj,Romano:2013kua,Romano:2012tt}.
More recently there has been some evidence that LTB solutions cannot provide a fully consistent cosmological model
compatible with all available observations \cite{Moss:2011ze,Zhang:2010fa}. These studies are nevertheless based on fitting experimental data with some particular functional ansatz for the functions defining the model, and as such they do not explore the full space of all possible inhomogeneity profiles.
The numerical inversion approach adopted in the present paper is on the contrary able to explore the full range of all the possible LTB solutions and initial conditions.

Since the inversion equations are numerically unstable around the redshift corresponding to the maximum of the angular diameter distance, a local Taylor expansion is necessary around that point, for which our fully analytical version of differential equations is particularly suitable.  Our approach provides a local Taylor expansion of the solution at any point, and the numerical solution of the differential equations is more stable since we don't need to integrate numerically the background equations. The use of the analytical solution allows also to set exactly the necessary initial conditions, while in previous attempts \cite{Chung:2006xh} it was necessary to derive some approximate consistency condition. As a result we can get a more accurate solution of the IP, and we are also able to explore the full class of LTB models with an arbitrary value of the central curvature in a self-consistent way.
We apply this method to invert the theoretical luminosity distance function corresponding to the best find $\Lambda CDM$ parameters and find that only a certain range of central of initial conditions is allowed, since for other models a transition from red to blue shift occurs, making these models incompatible with the observed luminosity distance.
We also show that the value of the Hubble parameter at the last scattering surface is independent of the central value of the curvature, and differ by about 20\% from the best fit $\Lambda CDM$ value as constrained by CMB observations. This  confirms the necessity to introduce a bang function to fit CMB data with LTB models, but contrary to previous numerical inversion studies it shows it independently of the central value of the functions defining the LTB model, by exploring the full class of possible initial conditions. 

The method we developed  does not need to be applied to LTB metrics as cosmological models describing the local universe around us, but could be applied to study the  effects of large scale inhomogeneities for a generic observer located inside some region of the Universe corresponding to a local oversensitivity or underdensity which cannot be modeled simply perturbation of a FLRW metric.

\section{Lemaitre-Tolman-Bondi (LTB) Solution\label{ltb}}
Lemaitre-Tolman-Bondi  solution can be
 written as \cite{Lemaitre:1933qe,Tolman:1934za,Bondi:1947av}
\begin{eqnarray}
\label{eq1} %
ds^2 = -dt^2  + \frac{\left(R,_{r}\right)^2 dr^2}{1 + 2\,E}+R^2
d\Omega^2 \, ,
\end{eqnarray}
where $R$ is a function of the time coordinate $t$ and the radial
coordinate $r$, $R=R(t,r)$, $E$ is an arbitrary function of $r$, $E=E(r)$
and $R,_{r}=\partial R/\partial r$.

Einstein's equations give
\begin{eqnarray}
\label{eq2} \left({\frac{\dot{R}}{R}}\right)^2&=&\frac{2
E(r)}{R^2}+\frac{2M(r)}{R^3} \, , \\
\label{eq3} \rho(t,r)&=&\frac{2 M,_{r}}{R^2 R,_{r}} \, ,
\end{eqnarray}
with $M=M(r)$ being an arbitrary function of $r$ and the dot denoting
the partial derivative with respect to $t$, $\dot{R}=\partial R(t,r)/\partial t$.
 The solution of Eq.\ (\ref{eq2}) can be expressed parametrically 
in terms of a time variable $\eta=\int^t dt'/R(t',r) \,$ as
\begin{eqnarray}
\label{eq4} \tilde{R}(\eta ,r) &=& \frac{M(r)}{- 2 E(r)}
     \left[ 1 - \cos \left(\sqrt{-2 E(r)} \eta \right) \right] \, ,\\
\label{eq5} t(\eta ,r) &=& \frac{M(r)}{- 2 E(r)}
     \left[ \eta -\frac{1}{\sqrt{-2 E(r)} } \sin \left(\sqrt{-2 E(r)}
     \eta \right) \right] + t_{b}(r) \, ,
\end{eqnarray}
where  $\tilde{R}$ has been introduced to make clear the distinction
 between the two functions $R(t,r)$ and $\tilde{R}(\eta,r)$
 which are trivially related by 
\begin{equation}
R(t,r)=\tilde{R}(\eta(t,r),r) \, ,
\label{Rtilde}
\end{equation}
and $t_{b}(r)$ is another arbitrary function of $r$, called the bang function,
which corresponds to the fact that big-bang/crunches can happen at different
times. This inhomogeneity of the location of the singularities is one of
the origins of the possible causal separation \cite{Romano:2006yc} between 
the central observer and the spatially averaged region for models
 with positive $a_D$. 

We introduce the variables
\begin{equation}
 a(t,r)=\frac{R(t,r)}{r},\quad k(r)=-\frac{2E(r)}{r^2},\quad
  \rho_0(r)=\frac{6M(r)}{r^3} \, ,
\end{equation}
so that  
the Einstein equations
(\ref{eq2}) and (\ref{eq3}) are written in a form 
similar to those for FLRW models,
\begin{equation}
\label{eq6} ds^2 =
-dt^2+a^2\left[\left(1+\frac{a,_{r}r}{a}\right)^2
    \frac{dr^2}{1-k(r)r^2}+r^2d\Omega_2^2\right] \, ,
\end{equation}
\begin{eqnarray}
\label{eq7} %
\left(\frac{\dot{a}}{a}\right)^2 &=&
-\frac{k(r)}{a^2}+\frac{\rho_0(r)}{3a^3} \, ,\\
\label{eq:LTB rho 2} %
\rho(t,r) &=& \frac{(\rho_0 r^3)_{, r}}{3 a^2 r^2 (ar)_{, r}} \, .
\end{eqnarray}
The solution of Eqs.\ (\ref{eq4}) and (\ref{eq5}) can now be written as
\begin{eqnarray}
\label{LTB soln2 R} \tilde{a}(\tilde{\eta},r) &=& \frac{\rho_0(r)}{6k(r)}
     \left[ 1 - \cos \left( \sqrt{k(r)} \, \tilde{\eta} \right) \right] \, ,\\
\label{LTB soln2 t} t(\tilde{\eta},r) &=& \frac{\rho_0(r)}{6k(r)}
     \left[ \tilde{\eta} -\frac{1}{\sqrt{k(r)}} \sin
     \left(\sqrt{k(r)} \, \tilde{\eta} \right) \right] + t_{b}(r) \, ,
\end{eqnarray}
where $\tilde{\eta} \equiv \eta\, r = \int^t dt'/a(t',r) \,$.

In the rest of paper we will use this last set of equations 
and drop the tilde to make the notation simpler.
Furthermore, without loss of generality, we may set 
the function $\rho_0(r)$ to be a constant,
 $\rho_0(r)=\rho_0=\mbox{constant}$.

\section {Geodesic equations}
The luminosity distance for an observer at the center of a LTB space 
as a function of the redshift is given by 
\be
D_L(z)=(1+z)^2 R\left(t(z),r(z)\right)
=(1+z)^2 r(z)a\left(\eta(z),r(z)\right) \,,
\ee
where $\Bigl(t(z),r(z)\Bigr)$ or $\Bigl((\eta(z),r(z)\Bigr)$
is the solution of the radial null geodesic equations.
The past-directed radial null geodesic is given by
\bea
\label{geo1}
\frac{dT(r)}{dr}=f(T(r),r) \,,
\quad
f(t,r)=\frac{-R_{,r}(t,r)}{\sqrt{1+2E(r)}} \,,
\eea
where $T(r)$ is the time coordinate along the geodesic as a function of the the coordinate $r$.
Applying the definition of red-shift it is possible to obtain \cite{Romano:2009xw}:
\bea
\label{geo3}
\frac{d \eta}{dz}
&=&\frac{\partial_r t(\eta,r)-F(\eta,r)}{(1+z)\partial_{\eta}F(\eta,r)}=p(\eta,r) \,,\\
\label{geo4}
\frac{dr}{dz}
&=&-\frac{a(\eta,r)}{(1+z)\partial_{\eta}F(\eta,r)}=q(\eta,r) \,. 
\eea
where we have used
\bea
f(t(\eta,r),r)&=&F(\eta,r) \,,\\
\dot{f}(t(\eta,r),r)&=&{1\over a}\partial_{\eta} F(\eta,r) \,,\\
R_{,r}(t,r)&=&\partial_r R(t(\eta,r),r)+\partial_{\eta} R(t(\eta,r),r) \partial_r \eta \,,\\
F(\eta,r)&=&-\frac{1}{\sqrt{1-k(r)r^2}}\left[\partial_r (a(\eta,r) r)
+\partial_{\eta} (a(\eta,r) r) \partial_r \eta\right]  \, \nonumber \\
&=&-\frac{1}{\sqrt{1-k(r)r^2}}\left[\partial_r (a(\eta,r) r)
-\partial_{\eta} (a(\eta,r) r) a(\eta,r)^{-1}\partial_r t \right]\,.
\eea

The functions $p,q,F$ have an explicit analytical form which can be obtained from $a(\eta,r)$ and $t(\eta,r)$.
Using this approach the coefficients of equations (\ref{geo3}) and (\ref{geo4}) are 
fully analytical, which is a significant improvement over previous 
methods which required a numerical integration of the Einstein's equations to obtain the function $R(t,r)$.
This version of the geodesics equations is suitable for both numerical and analytical applications, in particular will be useful to obtain 
a red-shift expansion of the inversion equations around the apparent horizon point.

\section{Initial conditions}

Before deriving the set of differential equations for the solution of the inversion problem it is important to analyze how many independent initial conditions we need to fix.
Our final goal will be to set and solve a set of differential equations in red-shift space starting from the center, where by definition $z=0$. Given our choice of coordinates the model will be fully determined
by the functions $k(z),r(z),\eta(z)$, corresponding to three initial conditions
\bea
r(0)   &=&0   \, \nonumber \\
\eta(0)&=&\eta_0  \, \nonumber \\
k(0)   &=&k_0     \,.
\eea
The system of differential equation we will derive only involves derivatives of order one respect to the red-shift, so these initial conditions will be enough.
Given the assumption of the central location of the observer we have $r_0=0$, while the observed value of the Hubble parameter $H_0$ corresponds to another constraint among the central values $k_0,\eta_0$, so only one of them is independent.
After defining the Hubble rate as 
\bea
H^{LTB}&=&\frac{\partial_t a(t,r)}{a(t,r)}=\frac{\partial_{\eta} a(\eta,r)}{a(\eta,r)^2} 
\eea
we need to impose the two following conditions 
\bea
a(\eta_0,0)&=&a_0 \,, \label{a0}\\
H^{LTB}(\eta_0,0)&=&H_0 \label{H0} ,
\eea
where $a_0$ is, as expected, an arbitrary parameter,  $\eta_0$ is the value of the generalized conformal time coordinate $\eta$ corresponding to the central observer today, and $H_0$ is the observed value of the Hubble parameter.

After re-writing the solution in terms of the following more convenient dimensionless quantities
\cite{Romano:2012ks}
\bea
a(T,r)&=&\frac{a_0 \Omega^0_M \sin ^2\left(\frac{1}{2} T \sqrt{K(r)}\right)}{K(r)} \,,\\
t(T,r) &=& H_0^{-1}\frac{\Omega^0_M}{2K(r)}
     \left[ T -\frac{1}{\sqrt{K(r)}} \sin
     \left(\sqrt{K(r)} \, T \right) \right] + t_{b}(r) \,,\\
k(r)&=&(a_0 H_0) K(r)\,, \\
\eta&=&T(a_0 H_0)^{-1}\,,\\
\rho_0&=&3 \Omega^0_m a_0^3 H_0^2\,.
\eea
we can solve eq.(\ref{a0},\ref{H0}) for $\Omega^0_M$ and $T_0$ to finally get the initial conditions and the exact solution in this form 
\bea
a(T,r)&=&\frac{a_0 (K_0+1) \sin ^2\left(\frac{1}{2} T \sqrt{K(r)}\right)}{K(r)} \,,\\
t(T,r) &=& H_0^{-1}\frac{1+K_0}{2K(r)}
     \left[ T -\frac{1}{\sqrt{K(r)}} \sin
     \left(\sqrt{K(r)} \, T \right) \right] + t_{b}(r) \,, \\
K_0&=&K(0) \,,  \\
T_0&=&\frac{\arctan{(2 \sqrt{K_0})}}{\sqrt{K_0}}\,\label{T0}\,\\
\Omega^0_m &=& K_0+1 \label{Om}\,.
\eea

Since we have three unknown $\{\Omega_M^0,T_0,K_0\}$ and two constraints given by eq.(\ref{a0},\ref{H0}) , one of them can always remain free, and the other two can be expressed in terms of it. In this paper we have chosen $K_0$ to be the free parameter, but we could equivalently chose another one.
The above form of the solution is particularly useful to explore the full class of LTB models since $K_0$ is a free parameter which determines through equation (\ref{T0}) the central value of the generalized dimensionless conformal time variable $T_0$.
$H_0$ is also a free parameter which can be set according to observations and fixes the scale for the definition of the dimensionless quantities ${K(r),T,\Omega_m^0}$.
This means that we can arbitrarily fix $K_0$ and $H_0$ as long as we impose the correct initial condition given by eq.(\ref{T0}-\ref{Om}). 

As expected $a_0$ does not appear in observable quantities such as the cosmic time $t(\eta,r)$, and it can be fixed to 1.
It can be easily checked that the above solution is by construction in agreement with any given value of $H_0$
\bea
H_0^{LTB}&=&\frac{\partial_t a(t_0,0)}{a(t_0,0)}=\frac{\partial_{\eta} a(\eta_0,0)}{a(\eta_0,0)^2}=(a_0 H_0)\frac{\partial_{T}a(T_0,0)}{a(T_0,0)^2}=H_0\,,
\eea
and for any $K_0$ we can now determine the corresponding initial condition $T_0=T(z=0)$.
In this way we can self-consistently determine all the necessary initial conditions and we are left with the freedom to fix $K_0$ arbitrarily.
As we will see later only some values of $K_0$ are consistent with observations. 
Our general approach to determine the initial conditions will allow us to explore the full class of LTB models, while in previous studies the value
of $K_0$ has been fixed \cite{Chung:2006xh}, and the initial conditions were based on some approximate consistency relation. 

\begin{figure}[h]
\includegraphics[width=7cm,height=7cm]{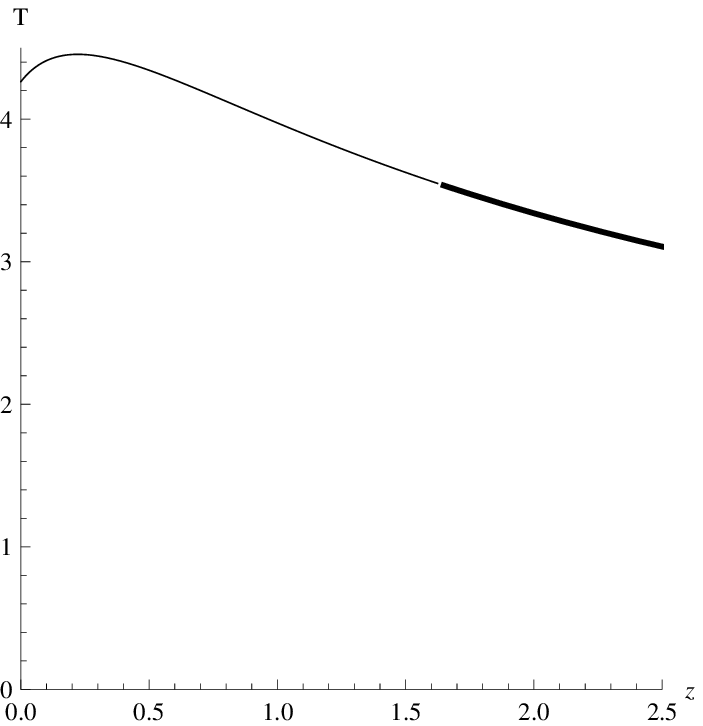}
\includegraphics[width=7cm,height=7cm]{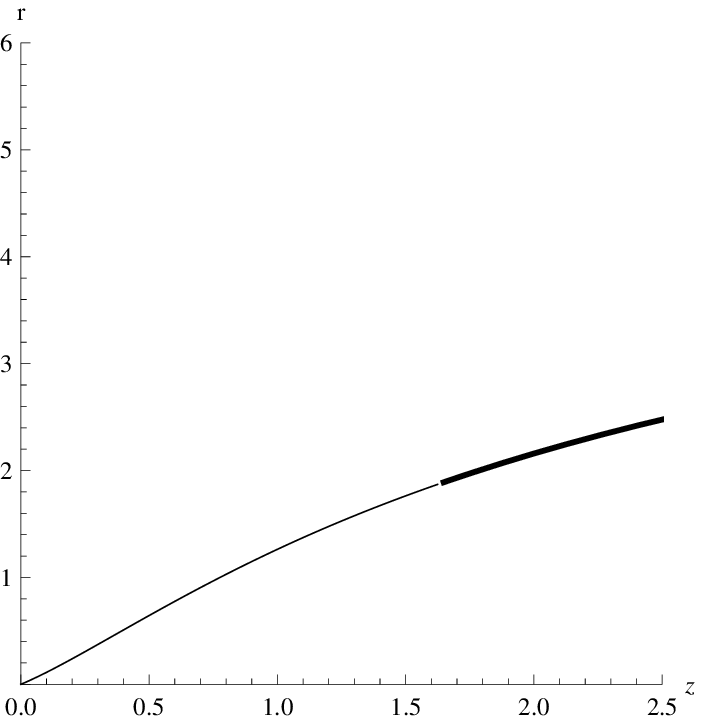}
\caption{The conformal time $T(z)$ and the radial coordinate $r(z)$ are plotted as a function of redshift for $K_0$=-0.9376. The thick line correspond to the part after the apparent horizon.}
\end{figure}

\begin{figure}[h]
\includegraphics[width=7cm,height=7cm]{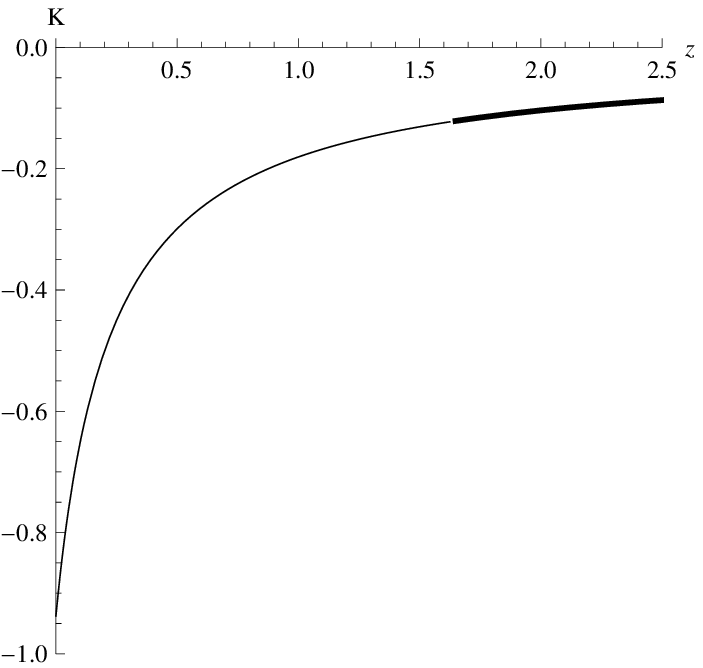}
\includegraphics[width=7cm,height=7cm]{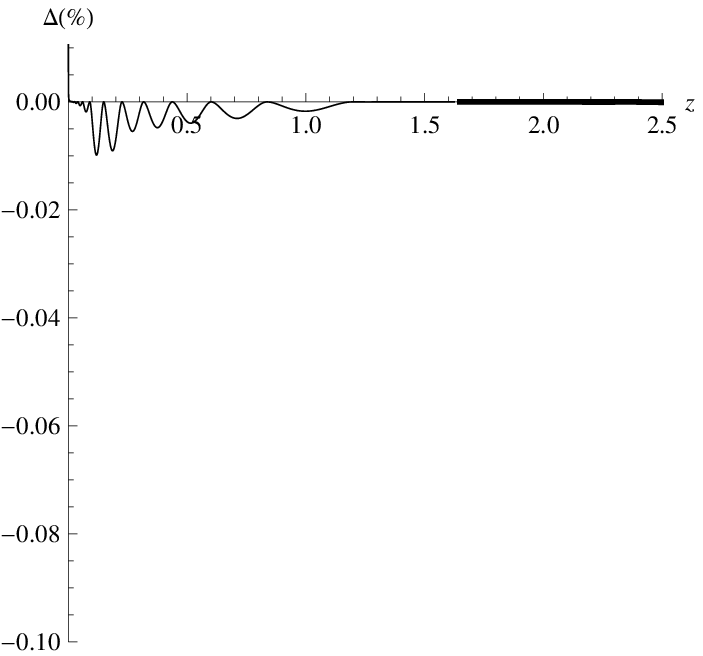}
\caption{The curvature function $K(z)$ and the relative percentual $\Delta(z)=100\frac{D^{\Lambda CDM}(z)-D^{LTB}(z)}{D^{\Lambda CDM}(z)}$ error between the luminosity distance $D^{\Lambda CDM}(z)$ used as input and $D_L^{LTB}(z)$ obtained by substituting the numerical solution of the differential equations for the inversion method are plotted as functions of redshift for $K_0$=-0.9376. The thick lines correspond to the part after the apparent horizon.}
\end{figure}

\begin{figure}[h]
\includegraphics[width=7cm,height=7cm]{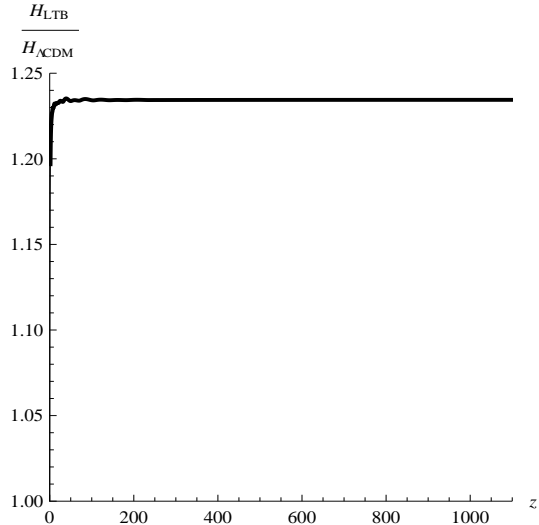}
\caption{The ratio between the Hubble parameter $H^{LTB}(z)$ and $H^{\Lambda CDM}(z)$ is plotted as function of the redshift for $K_0$=-0.9376. The thick lines correspond to the part after the apparent horizon.}
\end{figure}

\begin{figure}[h]
\includegraphics[width=7cm,height=7cm]{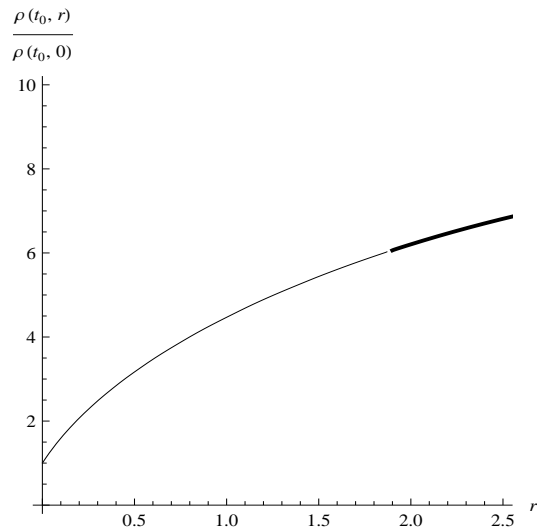}
\caption{The energy density profile $\frac{\rho \left(r,t_0\right)}{\rho \left(r\to 0,t_0\right)}$ is plotted as function of the comoving radius for $K_0$=-0.9376. The thick lines correspond to the part after the apparent horizon.}
\end{figure}

\begin{figure}[h]
\includegraphics[width=7cm,height=7cm]{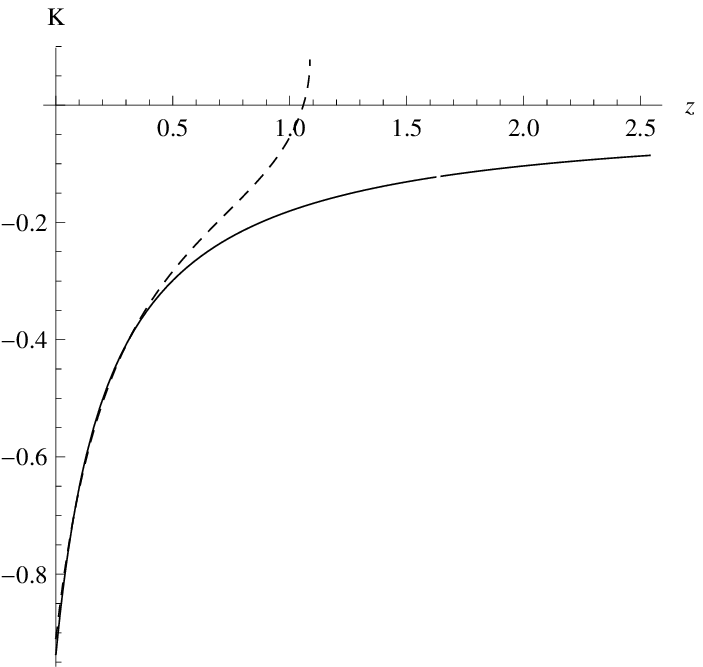}
\includegraphics[width=7cm,height=7cm]{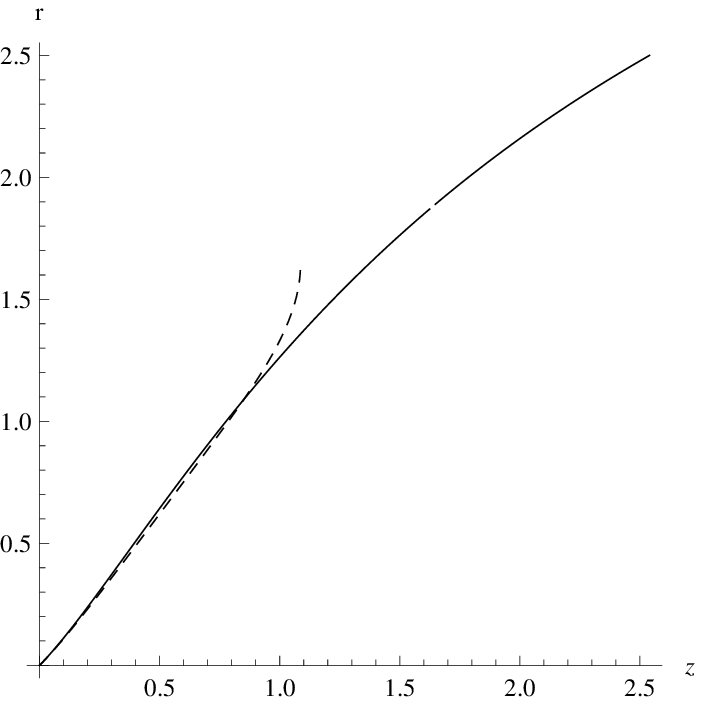}
\caption{The curvature $K(z)$ and the the comoving radial coordinate $r(z)$ are plotted as a function of the redshift. The black dashed line corresponds to $K_0=-0.91$ and the black line corresponds to $K_0=-0.9376$. As it can be seen for $K_0=-0.91$ there is critical redshift after which $r$ is decreasing, corresponding to a transition from red to blue shift. This implies that this choice of the initial conditions is not compatible with the observed expansion of the Universe.}
\end{figure}

\begin{figure}[h]
\includegraphics[width=7cm,height=7cm]{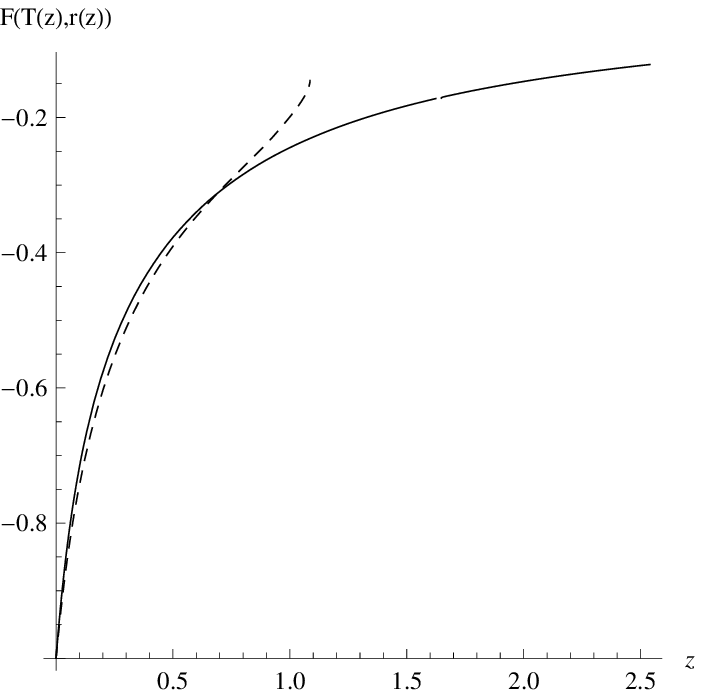}
\includegraphics[width=7cm,height=7cm]{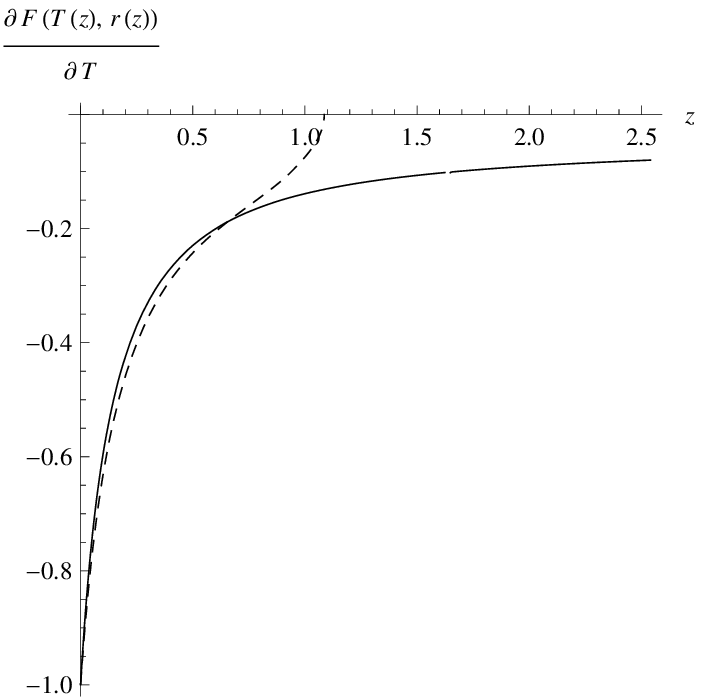}
\caption{$F(T(z),r(z))$ and $\frac{\partial F(T(z) ,r(z))}{\partial T }$ are plotted as functions of the red shift. Since $F(\eta,r) \propto \, \partial _rR(t(z),r(z))$ never crosses zero, no shell crossing singularity appears along the past light cone. The point where $\frac{\partial F(T(z) ,r(z))}{\partial T }$ crosses zero corresponds to the transition from local expansion to local contraction, i.e from red to blue shift. The  dashed line corresponds to $K_0=-0.91$ and the solid line corresponds to $K_0=-0.9376$. }
\end{figure}

\section{Inversion method differential equations}
In the previous section we have seen that it is possible to derive a fully analytical set of radial null geodesics equations. Our goal now is to use these equations to obtain a new set of differential equations to map an observed $D_L(z)$ to a LTB model.
In principle we need are three independent functions to fully specify a LTB solution, $M(r),k(r),t_b(r)$, but because of the freedom in fixing the radial coordinate only two of them are really independent. As explained in the previous sections we adopt the coordinates system in which $M(r)\propto r^3$, so that only $k(r),t_b(r)$ are left to be determined.
In this paper we will consider the case in which $t_b(r)=0$, since we are inverting only one observable, the luminosity distance $D_L(z)$, and the inversion method will be enough to fully determine the remaining function $k(z)$, and then $k(r)$. We will leave to a future work the case in which an additional redshift dependent observable is included, which will then require to also develop an inversion method for $t_b(r)$.
In the coordinates we have chosen a LTB solution is determined uniquely by the function $k(r)$, so we will have a total of three independent functions to solve for $\eta(z),r(z),k(z)$. Since we have already two differential equation for the geodesics, we need an extra differential equation.

This can be obtained by differentiating respect to the redshift the luminosity distance $D_L(z)$
\bea
\frac{d}{d z} \bigg(\frac{D^{obs}_L(z)}{(1+z)^2}\bigg)=\frac{\partial (r a(\eta,r))}{\partial \eta}\frac{d\eta}{d z}+\frac{\partial (r a(\eta,r))}{\partial r}\frac{d r}{d z}=s(z) \,
\eea
where $D^{obs}_L(z)$ is the  observed luminosity distance. In our case we will use the best fit $\Lambda CDM$ function.
Now we have the set of equations we were looking for
\bea
\frac{d \eta}{d z}=p(\eta(z),r(z))=p(z)\,,\label{E1}\\
\frac{d r}{d z}=q(\eta(z),r(z))=q(z)\,,\label{E2}\\
\frac{d}{d z} \bigg(\frac{D^{obs}_L(z)}{(1+z)^2}\bigg)=s(z)\,. \label{E3}
\eea
Since we will solve our differential equations respect to the the variable $z$, we need to transform the partial derivatives respect to $\eta$ and $r$ in eq.(\ref{geo3},\ref{geo4}) according to the chain rule:
\bea
\frac{\partial h(\eta,r)}{\partial r}\bigg|_{(\eta=\eta(z),r=r(z))}=\frac{\partial h(\eta(z),r(z))}{d z}\frac{d z}{d r} \,,\\ \label{ch1}
\frac{\partial h(\eta,r)}{\partial \eta}\bigg|_{(\eta=\eta(z),r=r(z))}=\frac{\partial h(\eta(z),r(z))}{d z}\frac{d z}{d \eta} \,. \label{ch2}
\eea
where $h(\eta,r)$ is a generic function in the coordinates $(\eta,r)$.
After this substitution the equations contain only functions of the red-shift $z$, and derivatives respect to $z$.
The differential equations obtained in this form need to be further manipulated in order to re-write them in a canonical form in which the derivatives appear all on one side, since after  the application of the chain rule to eq.(\ref{geo3},\ref{geo4}) derivative terms like $\frac{d r(z)}{d z},\frac{d \eta(z)}{d z},\frac{d k(z)}{d z}$ are also on the right-hand side.
We can now use eq.(\ref{ch1}-\ref{ch2}) in eq.(\ref{E1}-\ref{E3}) and after a rather complicated algebraic manipulation we get :
\bea
&& 2 t^2 \sqrt{K(z)} ((6+4 t^2) r(z)+(3+t^2) \sqrt{1-K(z) r(z)^2}T(z)) K'(Z)-12 t^3 \sqrt{1-K(z) r(z)^2} K'(Z)+ \nonumber \\
&& -8 t^3 (1+z) K(z)^2 r'(z) T'[z]-2 t K(z) r(z) K'(z) (3 (1+t^2)T(z)+(3+5 t^2) (1+z) T'(z))+\nonumber\\
&& +K(z)^{3/2} (-8 t^4 r'(z)+3 (1+t^2)^2 (1+z) r(z)T(z) K'(z) T'(z))=0 \label{e1} \\ 
&& r'(z) (2 t (3+5 t^2) (1+z) r(z) K'(z)-\sqrt{K(z)} (8 t^4 \sqrt{1-K(z) r(z)^2}+ \nonumber \\
&& +3 (1+t^2)^2 (1+z) r(z)T(z) K(z))+8 t^3 (1+z) K(z) r'(z))=0 \label{e2} \\ 
&& 2 K(z) ((1+K_0) t^2 r'(z)-(1+t^2) K(z) H_0 \frac{d}{d z} \bigg(\frac{D^{obs}_L(z)}{(1+z)^2}\bigg)) + \nonumber \\
&&-(1+K_0) t\, r(z) ((2 t-\sqrt{K(z)}T(z)) K(z)-2 K(z)^{3/2} T'(z))=0 \label{e3}
\eea
In the above expressions we have expressed all the trigonometric functions in terms of the equivalent expressions in terms of $\tan(X)$
according to
\bea
t&=&\tan(X)\,,\\
X&=&\frac{1}{2} \sqrt{K(z)} T (z)\,.
\eea
This is achieved by using a series of Mathetica simplifying routines developed for this purpose.
We have also used the dimensionless version of the solution in terms of $K(z),T(z)$ derived in the previous section.

As it can be seen the above three equations are not linear in the derivative terms, but the second one only involves $\{r'(z),K'(z)\}$, while the other two involve all the three functions $\{r'(z),K'(z),T'(z)\}$.
This suggests that we can first solve for $r'(z)$ in terms of only $K'(z)$ from the equation (\ref{e2}):
\bea
r'(z)&=&\frac{1}{8 t^3 (1+z) K(z)}\bigg[8 t^4 \sqrt{K(z)} \sqrt{1-K(z) r(z)^2}-6 t r(z) K'(z)-10 t^3 r(z) K'(z)-6 t\, z\, r(z) K'(z) \nonumber \\
&& -10 t^3 \,z\, r(z) K'(z)+3 \sqrt{K(z)} r(z)T(z) K'(z)+6 t^2 \sqrt{K(z)} r(z)T(z) K'(z)+ \nonumber \\ 
&&+ 3 t^4 \sqrt{K(z)} r(z) T(z) K'(z)+3 z \sqrt{K(z)} r(z) T(z) K'(z)+6 t^2 z \sqrt{K(z)} r(z) T(z) K'(z)+ \nonumber \\
&&+3 t^4\, z \sqrt{K(z)} r(z) T(z) K'(z)\bigg]
\eea
and then substitute in equations(\ref{e1},\ref{e3}) to get:
\bea
K'(z)&=&-\frac{1}{1+z}t (12 t^2 (1+z) \sqrt{1-K(z) r(z)^2} K(z)-2 t (1+z) \sqrt{K(z)} (9 (1+t^2) r(z)+ \nonumber \\
&&+(3+t^2) \sqrt{1-K(z) r(z)^2} T(z)) K(z)+K(z) (8 t^4 \sqrt{1-K(z) r(z)^2}+\nonumber \\
&& +3 (3+4 t^2+t^4) (1+z) r(z) T(z) K(z))+8 t^3 (1+z) K(z)^{3/2} \sqrt{1-K(z) r(z)^2}T'(z))\\
T'(z)&=&\frac{1}{4 t (1+z)}(-6 (1+K_0) t (1+3 t^2) (1+z) r(z) K(z)+(1+K_0) \sqrt{K(z)} (8 t^4 \sqrt{1-K(z) r(z)^2}+ \nonumber \\
&&+(3+10 t^2+3 t^4) (1+z) r(z) T(z) K(z))-8 t (1+t^2) (1+z) K(z)^2 H_0\frac{d}{d z} \bigg(\frac{D^{obs}_L(z)}{(1+z)^2}\bigg)+ \nonumber \\
&&8 (1+K_0) t^2 (1+z) K(z)^{3/2} r(z)T'(z))
\eea

These two equations now only involve ${K'(z),T'(z)}$ in a linear form, so they can be solved directly, and then the result for $K'(z)$ can be substituted in the equation for $r'(z)$. 
After some rather cumbersome algebraic manipulations we finally get:
\bea
\frac{d T(z)}{d z}&=&\frac{2 \sqrt{K(z)}}{3 t (1+K_0) r(z)} \times \Bigg[ H_0 \frac{d}{d z} \bigg(\frac{D^{obs}_L(z)}{(1+z)^2}\bigg) \nonumber \\
&& +\: \frac{H_0 \frac{d}{d z} \bigg(\frac{D^{obs}_L(z)}{(1+z)^2}\bigg) \left(1+3 t^2\right) \sqrt{K(z)} r(z)}{2 \left(\sqrt{K(z)} r(z)-t \sqrt{1-K(z) r(z)^2}\right)}-\frac{(1+K_0) t^3 \sqrt{1-K(z) r(z)^2}}{\left(1+t^2\right) (1+z) K(z)^{3/2}}\Bigg]\,,\\
\frac{d r(z)}{d z}&=&-\frac{\sqrt{1-K(z) r(z)^2}}{3\text{  }t \left(t^2 X-3 t+3 X\right)} \times \Bigg[ \frac{H_0 \frac{d}{d z} \bigg(\frac{D^{obs}_L(z)}{(1+z)^2}\bigg) K(z) \left(t \left(3+5 t^2\right)-3 \left(1+t^2\right)^2 X\right)}{(1+K_0)\left(-\sqrt{K(z)} r(z)+t \sqrt{1-K(z) r(z)^2}\right)}  \nonumber \\
&& +\: \frac{2\text{  }t^2 \left(2 t^3-3 t^2 X+3 t-3 X\right)}{\left(1+t^2\right) (1+z) \sqrt{K(z)}}\Bigg]\,,\\
\frac{d K(z)}{d z}&=&\frac{4 t^2 \sqrt{K(z)} \sqrt{1-K(z) r(z)^2}}{3 (1+K_0) \left(1+t^2\right) (1+z) r(z) \left(t^2 X-3 t+3 X\right)} \times \nonumber \\
&&\left[\frac{H_0 \frac{d}{d z} \bigg(\frac{D^{obs}_L(z)}{(1+z)^2}\bigg) \left(1+t^2\right) (1+z) K(z)^{3/2}}{-\sqrt{K(z)} r(z)+t \sqrt{1-K(z) r(z)^2}}-(1+K_0) t^2\right]\,.
\eea
where
\bea
t&=&\tan(X)\,,\\
X&=&\frac{1}{2} \sqrt{K(z)} T (z)\,.
\eea
and we have used the dimensionless version of the solution in terms of $K(z),T(z)$ derived in the previous section.
The main advantage of these equations is that they are fully analytical, while other versions require a numerical integration of the Einstein's equations. In this form the central value both $H_0$ and $K_0$ can fixed arbitrarily, and the remaining initial condition for $T(z)$ are fixed according to eq.(\ref{T0}).  This makes them suitable both for numerical and analytical applications.
In particular they can be used to expand locally the solutions around the apparent horizon corresponding to the maximum of $D_A(z)=\frac{D_L(z)}{(1+z)^2}$.
We will report in the appendix the relations which can be used to obtain such an expansion.

\section {Apparent horizon, $H(z)$ and CMB}
As it can be easily seen the differential equations we need to solve become unstable around a critical vale of the redshift $z_c$, where the angular diameter distance 
\bea
D^{\Lambda CDM}_A(z)&=&\frac{1}{(1+z)^2} D^{\Lambda CDM}_L(z)
\eea
which we use as input for our differential equations reaches its maximum. This is only an apparent horizon, due to the fact that we are the using red-shift as the variable of the differential equations, not to a real singularity of the space-time.
In order to overcame this critical point we follow these steps :
\begin{itemize}
\item
Choose a point $z_c-\epsilon_1$ before the apparent horizon where the numerical inversion method is still sufficiently accurate and stable, and taking advantage of the fully analytical system of differential equations, Taylor expand $T(z),r(z),K(z)$.
\item Extrapolate the obtained Taylor expansions to a point after $z_c+\epsilon_2$ after the apparent horizon, sufficiently far to avoid the numerical instability and minimize the relative error between the extrapolated $D^{LTB}_{Taylor}(z)$ and $D^{Obs}_L(z)$.
\item Use the extrapolated values at  $z_c+\epsilon_2$ as the initial conditions for the numerical solution of the system of differential equations after the apparent horizon. 
\end{itemize}

This method is quite effective, as it can be seen in the plot of the relative error in Fig.(2), and allows to obtain a very accurate solution up to very high redshift.
We can get significantly more accurate results than previous ones \cite{Chung:2006xh},  even after the critical point, since the fully analytical expression of the equations we use allows to obtain a very accurate Taylor expansion near the critical point.

\section{Application : blue to red-shift transition and $H_0(z_{LSS})$ }
As it can be seen in the figures the inversion procedure is quite accurate, since we can keep the relative error between the solution of the inversion problem and $D_L^{obs}(z)$ quite low, much better than in \cite{Chung:2006xh}. This is due to fact that the initial conditions we are setting are exact while in previous studies they were only approximate.
Compared to \cite{Celerier:2009sv} this method is more accurate because we expand in red-shift space the actual geodesics equation around the critical point.

We can now apply the inversion method we have derived. We find that only certain values of $K_0$
allow to solve the differential equations up to high red-shift. For  sufficiently large $K_0$ we find in fact that the geodesics equations became unstable because we approach a point along the light cone where 
\bea
\frac{d z}{d r}&=&0
\eea 
i.e. there is a turning point from red-shift to blue-shift. This implies that these models are inconsistent with observations.
As it can be seen from the geodesics equations this can happen when:
\bea
\partial_{\eta} F(\eta,r)&=&0 \,,
\eea 
which in the $(t,r)$ coordinates is equivalent to 
\bea
\dot{R'}(t,r)=0 \,.
\eea
As shown in fig. (6) this is exactly what occurs for certain values of $K_0$, where we can also see that this is not a shell-crossing singularity, since $F(\eta,r)$ never crosses zero before that point.  
We can easily interpret this result using our intuition about the Friedman like equation in which the Einstein's equations can be written for the $LTB$ solution.
The curvature term has to be negative in order to mimic the effects of a cosmological constant, and if the central value is not sufficiently large than there can be some critical point where the matter gravitational attraction will dominate and cause a contraction.

Another important observable to fit is the CMB spectrum. Since the CMB physics is determined by $H^{LTB}(z_{LSS})$ it would be interesting to explore the possibility that an appropriate choice of $K_0$ could also give a good agreement between $H^{LTB}(z_{LSS})$ and $H^{\Lambda CDM}(z_{LSS})$ .
The numerical solution of the inversion problem shows that $H^{LTB}(z_{LSS})$ is not affected significantly by $K_0$ and that a mismatch of the order of the $20\%$ cannot be avoided, independently of the value of $K_0$. 
This implies than even taking into account the freedom on the choice of $K_0$ we cannot find any model such that
\bea
D^{LTB}_L(z)&=&D_L^{\Lambda CDM} \,,\\
H^{LTB}(z_{LSS})&=&H^{\Lambda CDM}(z_{LSS}) \,.
\eea
We deduce that none of these models should be able fit both $D_L(z)$ and the CMB spectrum, and it would be necessary the introduction of an extra functional degree of freedom, the bang function $t_b(r)$, to achieve that goal.
Since we have explored all the possible set of initial conditions for $K(z)$, it should be noted that our conclusion is more general than previous ones based on particular choice of $K_0$.

The reason is that the $K(z)$ solution is asymptotically constant (zero in our case since we mimic a flat FLRW model) because at sufficiently high redshift, where the cosmological constant is subdominant, the homogenous FLRW Universe has to be recovered. This implies that the low redshift disagreement between $H^{LTB}$ and $H^{FLRW}$ remain the same at high redshift, and in general they don't intersect, as long as we keep solving the inversion problem for $D_L(z)$ at any red-shift. 


\section{Conclusions}
We have developed a new fully analytical inversion method to map the observed luminosity distance to a LTB model.
This method has the advantage of not requiring any numerical integration of the Einstein's equations, and is particularly suitable to obtain a Taylor expansion of the solution around the numerical instability point corresponding to the apparent horizon.
The accuracy of the solution we obtain significantly improves previous methods, because we are able to  fix exactly initial conditions and the Taylor expansion in red-shift is very precise, allowing to overcame the apparent horizon keeping the relative error low.

We have tested this inversion method to investigate the importance of the choice of the initial central value $K_0$ for the curvature function defining the LTB model. We found that only a certain range of values is consistent with the observed cosmic red-shift, since higher values of $K_0$ lead to a transition from red to blue shift.
We have also checked that the high redshift value of $H^{LTB}$ is not affected significantly by $K_0$, and that all the acceptable models, i.e. the ones without blue to red-shift transition, have a disagreement of the order of $20\%$ respect to $H^{\Lambda CDM}(z_{LSS})$.
Since we have explored all the possible set of initial conditions for $K(z)$, it should be noted that our conclusion is more general than previous ones based on a particular choice of $K_0$ or ansatz for some of the functions defining the LTB model.
In the future it will be interesting to extend this method to the case of a not vanishing bang function $t_b(r)$ in order to solve the inversion problem also for the Hubble parameter as a function for the red-shift.

The method we developed  does not need to be applied to LTB metrics as cosmological models describing the local universe around us, but could be applied to study the effects of large scale inhomogeneities for a generic observer located inside some region of the Universe corresponding to a local oversensitivity or underdensity which cannot be modeled simply perturbation of a FLRW metric.

\begin{acknowledgments}
Chen and Romano are supported by the Taiwan NSC under Project No.\
NSC97-2112-M-002-026-MY3, by Taiwan's National Center for
Theoretical Sciences (NCTS). Chen is also supported by the US Department of Energy
under Contract No.\ DE-AC03-76SF00515.
Romano also thanks the Perimeter Institute for its hospitality.
\end{acknowledgments}

\appendix
\section{Expansion around the apparent horizon}
In this appendix we give an explicit form of the coefficients of the expansion of the geodesics equation around the apparent horizon $z_c$. The linear coefficients  are simply given by evaluating the right hand side of the geodesics equations at $z_c$. Since the right hand side of the geodesics equations is fully analytical, we can take its first derivative, and then solve for the second derivative terms to obtain:
\bea
\frac{d^2K(z)}{dz^2}&&\frac{d\ln K(z)}{dz}\Bigg[\frac{d\ln \left(1-K(z) r(z)^2\right)}{2 K(z) r(z)^2 dz}+\frac{d\ln K(z)}{dz}-\frac{1}{1+z}+\frac{d\ln \left(K(z)r(z)^2\right)}{dz}\times \nonumber \\
&& \left(\frac{2 X}{t}+\frac{2 t (t-X) X}{3t-\left(3+t^2\right)X}\right)+\frac{H_0 \frac{d^2}{dz^2}\left(\frac{D_L^{obs}(z)}{(1+z)^2}\right) \left(1+t^2\right) (1+z) K(z)}{\left(1+K_0\right) t^2 \left(1-t\sqrt{K(z)^{-1} r(z)^{-2}-1}\right)r(z)}\Bigg]\,,\\
\frac{d^2r(z)}{dz^2}&=&\frac{d\ln  r(z)}{dz}\Bigg[\frac{d\ln \left(1-K(z) r(z)^2\right)}{2 K(z) r(z)^2 dz}+\frac{d\ln  r(z)}{dz}-\frac{1}{1+z}+\frac{X d\ln \left(K(z)r(z)^2\right)}{3 dz}\times \nonumber \\
&& \left(\frac{3+t^2}{2t}+\frac{2 t^3 X}{3 t-\left(3+t^2\right) X}+\frac{3 t^4}{t \left(3+2t^2\right)-3 \left(1+t^2\right) X}\right)-\frac{H_0 (1+z)}{1+K_0}\times \nonumber \\
&& \left.\frac{\frac{d^2}{dz^2}\left(\frac{D_L^{\text{obs}}(z)}{(1+z)^2}\right) \left(1+t^2\right) K(z)}{t^2 \left(1-t\sqrt{K(z)^{-1} r(z)^{-2}-1}\right) r(z)}\left(1+\frac{3t^3-3 \left(1+t^2\right)t^2 X }{t \left(3+2t^2\right)-3\left(1+t^2\right)X}\right)\right] \,,\\
\frac{d^2T(z)}{dz^2}&=&\frac{d\ln T(z)}{dz}\Bigg[\frac{d\ln \left(1-K(z) r(z)^2\right)}{2 K(z) r(z)^2 dz}-\frac{d\ln K(z)}{2dz}-\frac{1}{1+z}+\frac{X d\ln \left(K(z)r(z)^2\right)}{t dz} \nonumber \\
&& -\: \frac{H_0 \frac{d^2}{dz^2}\left(\frac{D_L^{obs}(z)}{(1+z)^2}\right) \left(1+t^2\right) (1+z) K(z)^{3/2}}{2\left(1+K_0\right) t^3 \sqrt{1-K(z) r(z)^2}}\bigg(2+\frac{1+3t^2}{1-t \sqrt{K(z)^{-1} r(z)^{-2}-1}}\bigg)\Bigg]\,.
\eea
The derivation of the these coefficients is taking into account that $D_A'(z_c)=0$ and involves a series of cumbersome  algebraic and trigonometric manipulations of the type used in the derivation of the inversion equations, which have been carried out using a set of routines written in MATHEMATICA for this specific purpose.

\end{document}